\documentstyle[twoside,fleqn,espcrc2,epsfig]{article}

\def\beq{\begin{equation}}
\def\eeq{\end{equation}}
\def\bea{\begin{eqnarray}}
\def\eea{\end{eqnarray}}

\newcommand{\beqn}{\begin{eqnarray}}
\newcommand{\eeqn}{\end{eqnarray}}

\title{
\vspace{-9mm}
\rightline{\small ITEP-LAT/2002-14}
\vspace{-2mm}
\rightline{\small {\bf 5} September, 2002}
Geometry of percolating monopole clusters
\thanks{Talk presented by P.Yu.~B. at Lattice 2002, Boston}}
\author{P.~Yu.~Boyko\address[ITEP]{Institute of Theoretical and Experimental
Physics, B.Cheremushkinskaya 25, Moscow, 117259, Russia},
M.I.~Polikarpov\addressmark[ITEP],
V.I.~Zakharov\addressmark[ITEP]$^{,}$\address[MP]{Max-Planck Institut f\"ur
Physik, F\"ohringer Ring 6, 80805 M\"unchen, Germany}}

\begin{document}

\begin{abstract}
We perform detailed measurements of the
geometrical characteristics of the percolating cluster of the
magnetic monopole currents in the confining phase of the lattice $SU(2)$
gluodynamics.
The Maximal Abelian projection is used to define the
monopoles. The use of the geometrical language is motivated
by recent observations that the full non-Abelian action
associated with the monopoles corresponds to point-like
particles on the currently available lattices.
Scaling behavior of various quantities is observed.  \end{abstract}

\maketitle

\section{INTRODUCTION}

The monopole condensation is one of the most favored mechanisms
of the confinement, for review and references see \cite{review}.
Still, there remain important unresolved questions concerning this
mechanism. The main point is that the definition of the monopoles,
which are Abelian in nature, is not unique within a non-Abelian
theory. From pure phenomenological point of view the monopoles
defined within the Maximal Abelian Projection (MAP) seem to be
most successful.

Monopoles within a particular gauge would not have been of
much use if it were not so that they possess remarkable
$SU(2)$ invariant properties. In particular, the density of monopoles in
the percolating cluster scales as a physical quantity of dimension
$d=3$ (see \cite{bornyakov} and references therein).
It is worth emphasizing that apart from the
percolating cluster that fills in the whole of the lattice
\cite{ivanenko} there exist also finite-size clusters. The percolating
cluster is responsible for the confinement and we will concentrate
here on its properties.

A further and dramatic evidence for the reality of the MAP monopoles
comes from measurements of the full, non-Abelian action associated
with the monopoles, see \cite{anatomy} and references therein.
Namely, it turned out that at presently available lattices
the action is large and ultraviolet divergent:
\beq\label{uv}
S_{mon}~=~c_{action}\cdot(L/a)\, ,
\eeq
where $c_{action}$ is a constant, $L$ is the
length of the monopole trajectory and $a$
is the lattice spacing. Moreover, the $const$ in Eq. (\ref{uv})
is close to $\ln 7$ which determines the entropy factor.
It is well known that the physical mass, entering the propagator
is proportional to the difference
\beq\label{fine}
m^2_{phys}\cdot a~\approx~{\big(c_{action}-\ln 7\big)/a}\, .
\eeq
The two terms in the r.h.s. of Eq. (\ref{fine})
cancel each other to a great extent and
one can say that monopoles are fine-tuned \cite{vz}.

These observation imply that at presently available lattices the
monopoles appear not as extended field configurations
but rather as geometrical,
point-like objects. Moreover, the monopoles are always observed as (closed)
trajectory. Therefore the use of the language of the polymer approach
to field theory, or of quantum geometry (see, e.g., \cite{ambjorn})
is quite natural.

Below, we present results of measurements of the geometrical
characteristics of the monopole trajectories belonging to the
percolating cluster. In more detail, the percolating cluster consists
of segments of trajectories connecting crossings. We measure,
in particular,
the average length of the segments and average Euclidean distance
between the crossings, the average value of the crossings
per unit of 4-volume. In all these cases we observe simple
scaling properties which confirm the reality of the MAP monopoles
as of physical objects.

\section{MEASUREMENTS}

\begin{table*}[htb]
\caption{$SU(2)$ configurations.}
\label{table:1}
\newcommand{\m}{\hphantom{$-$}}
\newcommand{\cc}[1]{\multicolumn{1}{c}{#1}}
\renewcommand{\tabcolsep}{1.37pc} 
\renewcommand{\arraystretch}{1.2} 
\begin{tabular}{l l l l l l l l l}
\hline
$\beta$ & 2.30 & 2.35 & 2.40 & 2.40 & 2.45 & 2.50 & 2.55 & 2.60 \\
$L$ & 16 & 16 & 16 & 24 & 24 & 24 & 28 & 28 \\
$N$ & 100 & 100 & 100 & 20 & 10 & 50 & 40 & 50 \\
\hline
\end{tabular}
\end{table*}

We study the geometry and scaling properties of the monopole trajectories
obtained in the maximal Abelian projection of $SU(2)$ lattice gauge theory.
To fix the abelian projection we use the simulated annealing algorithm
\cite{SA}. We study monopoles on symmetric, $L^4$, lattices ($16\le L\le
24$) for 7 values of $\beta$. The size of the lattice, $L$, the
coupling, $\beta$, and the number of the independent gauge field
configurations, $N$, is listed in Table 1.  To fix the physical scale we use
$\sqrt \sigma  = 440 \, Mev$.

In Section 3 we study the geometrical properties of the monopole
trajectories, in Section 4 we discuss the scaling properties of various
physical quantities related to monopole trajectories.

\section{LONG-RANGE CORRELATION}

The monopole trajectory is constructed from the links on the dual lattice.
It is made from segments (the trajectories between crossings) and crossings
of the trajectories.
Now we show that the monopole trajectory is not a random walk. Consider the
correlation of the direction of the link $C_1$ which form the monopole
current with the direction of some ``initial'' link $C_0$. The considered
link is at the distance $l$ along the monopole trajectory from the initial
link.  The direction $C_1$ can be the same as $C_0$, the opposite, or other
(neither the same nor the opposite). In our normalization all three
correlations are equal to unity for random walk if $l\neq 0$ (for $l=0$ the
opposite direction is forbidden). In Fig.~1 we show the correlations of the
directions for $\beta=2.6$ on the lattice $28^4$.

\begin{figure}
\caption{Correlations of the directions of the links lying on the
monopole current.}
\includegraphics[width=0.5\textwidth]{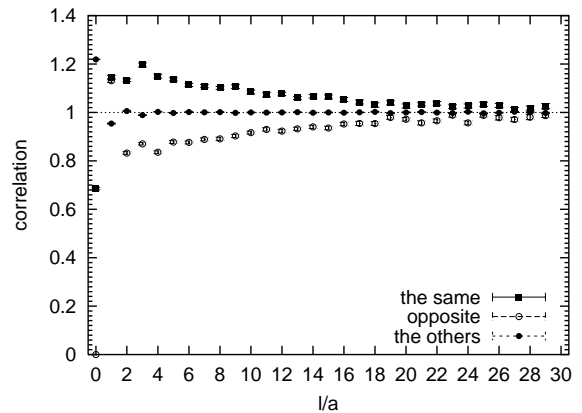}
\end{figure}

It is seen that the monopole trajectory has a "long memory". Even for links
separated by 17 lattice steps from the initial link the most probable
direction is the same as the direction of the initial link.

\section{SCALING PROPERTIES}

The average length of the segment of the monopole trajectory between
crossings $<l>$  scales, see from Fig.~2.

\vspace{4mm}
{Figure 2. $<l>$ vs. lattice spacing.}
\centerline{\includegraphics[width=0.5\textwidth]{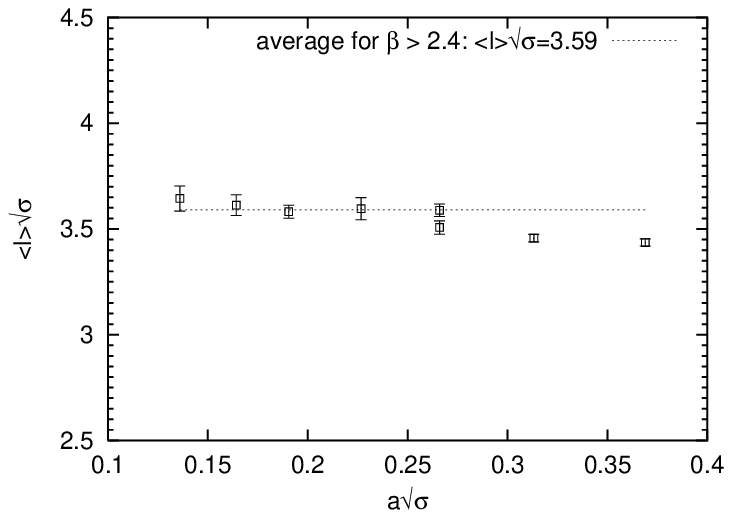}}

\vspace{4mm}
{Figure 3.~$<d>$ vs. lattice spacing.}
\centerline{\includegraphics[width=0.5\textwidth]{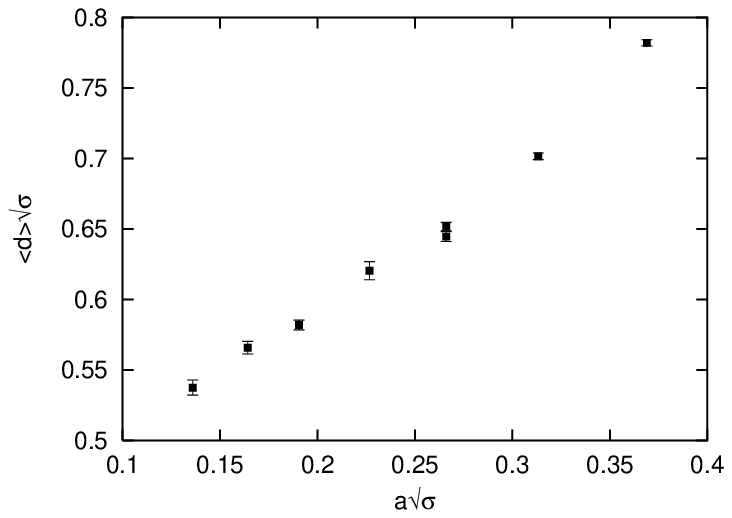}}

Numerically, $<l> \approx 1.6
\,\, fm$ if we average the data for
$\beta > 2.4$. This scaling behavior confirms strongly the fine tuning of
the action and entropy factors, see discussion following Eq. (\ref{fine}).

The data on the average Euclidean distance $<~d~>$ between crossings are
presented in Fig.~3.

The violations of the scaling are much more substantial
in this case. Still they can be approximated by a correction linear in the
lattice spacing $a$.

\vspace{4mm}
{Figure 4. Number of crossings per unit physical volume vs.
lattice spacing.}
\centerline{\includegraphics[width=0.5\textwidth]{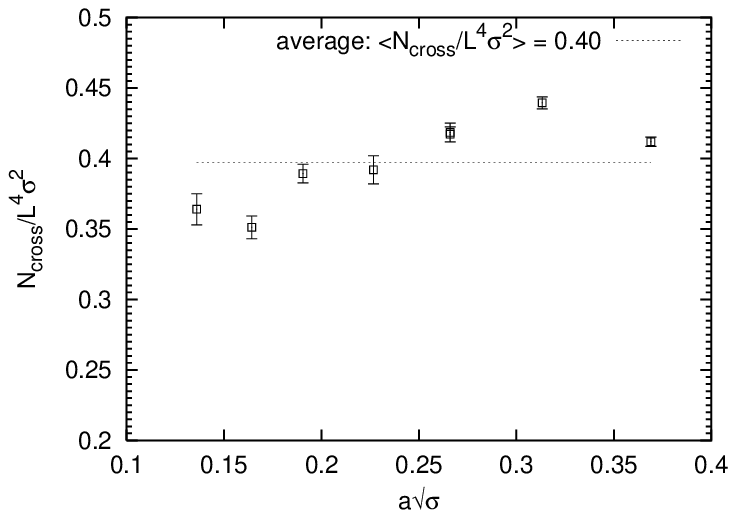}}

Finally, we found that the number of the crossings of the monopole
trajectories per unit physical volume seems to scale as it is seen from
Fig.~4.  The average $<N_{cross}>/L^4\sigma^2 \approx 0.4$ corresponds to
approximately 10 crossings per hypercube $1 \,\, fm^4$. This quantity is
probably related to $\phi^4$ constant in the effective monopole lagrangian,
$\phi$ is the monopole field.

\section*{ACKNOWLEDGEMENTS}
The authors are grateful to V.~G.~Bornyakov, M.~N.~Chernodub and
F.~V.~Gubarev for useful discussions.  M.~I.~P. is partially supported
by grants RFBR 02-02-17308, RFBR 01-02-117456, RFBR 00-15-96-786,
INTAS-00-00111, and CRDF award RPI-2364-MO-02. P.~Yu.~B. is partially
supported by grants RFBR 02-02-17308 and CDRF award MO-011-0. V.I.Z. is
partially supported by grant INTAS-00-00111 and DFG program "From lattice to
hadron phenomenology".

\end{document}